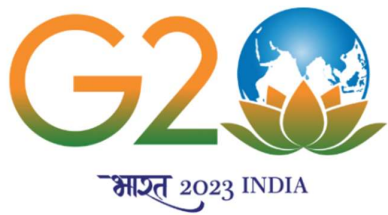 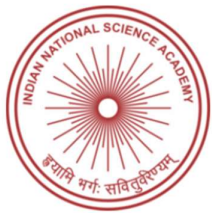 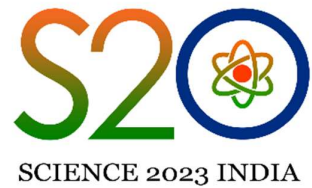

# Policy Brief

ASTROINFORMATICS: RECOMMENDATIONS FOR GLOBAL COOPERATION

Discussed and drafted during S20 Policy Webinar on Astroinformatics for Sustainable Development held on 6-7 July 2023

Contributors: Ashish Mahabal, Pranav Sharma, Rana Adhikari, Mark Allen, Stefano Andreon, Varun Bhalerao, Federica Bianco, Anthony Brown, S. Bradley Cenko, Paula Coehlo, Jeffery Cooke, Daniel Crichton, Chenzhou Cui, Reinaldo de Carvalho, Richard Doyle, Laurent Eyer, Bernard Fanaroff, Christopher Fluke, Francisco Forster, Kevin Govender, Matthew J. Graham, Renée Hložek, Puji Irawati, Ajit Kembhavi, Juna Kollmeier, Alberto Krone-Martins, Shri Kulkarni, Giuseppe Longo, Vanessa McBride, Jess McIver, Sanjit Mitra, Timo Prusti, A. N. Ramaprakash, Eswar Reddy, David H. Reitze, Reinaldo R. Rosa, Rafael Santos, Kazuhiro Sekiguchi, Kartik Sheth, Seetha Somasundaram, Tarun Souradeep, R. Srianand, Annapurni Subramaniam, Alex Szalay, Shriharsh Tendulkar, Laura Trouille, Yogesh Wadadekar, Patricia Whitelock

**Introduction**
In recent decades, the field of astronomy has undergone significant transformations, transitioning from the utilization of photographic plates to employing advanced technology like cameras equipped with charged coupled devices (CCDs). This pivotal shift, accompanied by enhanced storage capacities, accelerated processors, and improved communication capabilities, has ushered in an era characterized by comprehensive all-sky surveys and expansive international initiatives. The sheer magnitude of astronomical data generated by these endeavors has made manual analysis unfeasible.

In response to this data deluge, astronomers have collaborated with computer scientists and statisticians to harness the power of machine learning and other sophisticated methodologies. These collaborative efforts have culminated in the emergence of a new interdisciplinary field known as "astroinformatics." Astroinformatics holds tremendous promise in reshaping our comprehension of the universe and illuminating novel insights across a diverse spectrum of scientific domains.

Specifically, the methodologies originally devised for the analysis of astronomical data have already been successfully employed in and hold further potential for application in the domains of medical imaging and diagnostics, remote sensing, environmental monitoring, and sustainable economic practices. This cross-domain knowledge transfer facilitated by astroinformatics has the potential to contribute meaningfully to the resolution of critical global issues.

From July 6th to July 7th, 2023, a gathering of approximately 50 global experts convened to engage in comprehensive discussions pertaining to astroinformatics. The deliberations encompassed a wide array of topics, including broad astroinformatics, sky surveys, large-scale international initiatives, global data repositories, space-related data, regional and international collaborative efforts, as well as workforce development within the field. These discussions comprehensively addressed the current status, notable achievements, and the manifold challenges that the field of astroinformatics currently confronts.

The convergence of substantial data volumes with advanced data science capabilities underscores the pivotal role of data-driven scientific endeavors within this domain. It is worth noting that many investigations continue to commence with more conventional inquiries, often targeting readily available and easily identifiable individual celestial objects rather than striving to harness the full potential inherent in the vast reservoirs of collected data. This approach falls short of unlocking the complete scope of insights that these extensive data repositories have the potential to yield.

The G20 nations present a unique opportunity due to their abundant human and technological capabilities, coupled with their widespread geographical representation. Leveraging these strengths, significant strides can be made in various domains. These include, but are not limited to, the advancement of STEM education and workforce development, the promotion of equitable resource utilization, and contributions to fields such as Earth Science and Climate Science.

In the forthcoming sections, we present a concise overview, followed by specific recommendations that pertain to both ground-based and space data initiatives. Our team remains readily available to furnish further elaboration on any of these proposals as required. Furthermore, we anticipate further engagement during the upcoming G20 presidencies in Brazil (2024) and South Africa (2025) to ensure the continued discussion and realization of these objectives.

*Sky Surveys*: Sky surveys have been a crucial tool in advancing our understanding of the Universe. They have helped us discover new classes of astronomical objects and contribute significantly to our understanding of the origin of the elements, dark matter and dark energy, the accelerated expansion of the universe, and gravitational waves. They have helped us study the distribution of neutral and ionized matter in the Universe and test our theories about the origin and evolution of galaxies, stars, and planets. The search for Near-Earth objects has been a motivation for many surveys, and their cataloguing is important to secure humanity's continued existence. The last few decades have seen an explosion in the number and scope of sky surveys, both ground-based and space-based.

*Large projects - Cross Country Collaborations*: Many of the cutting-edge astronomy projects are extremely expensive and talent intensive. The ground based next generation optical and IR telescope facilities (e.g. GMT, TMT, and E-ELT), SKA for radio astronomy, and LIGO for gravitational astronomy, Gaia from space are a few examples of multi-billion-dollar projects. These have great potential for transformational scientific breakthroughs, technological innovation and advancing human knowledge about the universe.

Large projects, such as Gaia, LIGO, TMT, SKA, and others, serve as catalysts for cross-disciplinary collaboration, international partnerships, and the development of advanced data analysis techniques. By harnessing the power of astroinformatics, these projects are not only advancing our understanding of the cosmos but also contributing to sustainable development goals by promoting global collaboration, education, and technological advancements. Importantly, the large multinational collaborations become catalysts for fostering unity and peace among the global population as they have shared interest in the projects and their outcomes.

*Space Data*: The usage of computing accelerators such as GPUs (Graphics Processing Units) or FPGAs (Field Programmable Gate Arrays), are both good options to shorten processing times. The cost-benefit balance needs to be evaluated. Computation close to the data is in general good.

*Global Data*: The upcoming Vera C. Rubin Observatory[1], which will collect 20 TB of data per day and will optically image an estimated 38 billion objects over ten years, and the ambitious Square Kilometer Array project, which will collect an exabyte of raw radio data per day, compressed down to 10 PB/day, will generate more data than all previous telescopes combined. Processing, moving and curating these data will present one of the biggest challenges for the coming decades. The current Zwicky Transient Facility (ZTF[2]), which covers the night sky once every two days, is already demonstrating some of the complexities required in serving and processing "alerts" of new objects detected during the survey.

As such, astronomy has faced the issue of bespoke data formatting and data type conversion for many years. In the early 2000s, a global movement towards consistent data formats and standards grew to form the International Virtual Observatory Alliance (IVOA), with member consortia including the US Virtual Astronomical Observatory (VAO) which grew out of the US National Virtual Observatory (NVO), NASA's Planetary Data System (PDS), the Astronomical Virtual Observatory (from the European Southern Observatory or ESO), Astronet, the UK AstroGrid and more[3].

---

[1] Ivezić, Z et al. LSST: From Science Drivers to Reference Design and Anticipated Data Products ApJ 873 111 (2019)
[2] Bellm, E. et al The Zwicky Transient Facility: System Overview, Performance, and First Results PASP 131 018002 (2019)
[3] R. Hanisch et al. The Virtual Astronomical Observatory: Re-engineering access to astronomical data Volume 11, Part B, June 2015, Pages 190-209

The FAIR principles (that data should be Findable, Accessible, Interoperable, and Reusable)[4] are particularly valuable when applied in the astronomical context. It must be stressed that policies that enable and facilitate data sharing are the key to progress in this area, and that Astronomy can be a great way to test new ideas/models for sharing data given the relative openness of astronomical data. The G20 countries can share best practices for open access of code and make resolutions to adhere to FAIR principles.

*Workforce Development*: Currently the demographics in the field of astronomy and astroinformatics do not reflect the population in most nations.[5] We often find that certain groups are excluded from the field of astronomy and astrophysics.[6] In several countries, groups from rural areas, economically disadvantaged households and other marginalized groups are often left out of the astronomy and astrophysics ecosystem.

The International Astronomical Union (IAU) Office of Astronomy for Development (OAD) plays a significant role globally in assisting countries in achieving objectives like the UN's Sustainable Development Goal 8 (SDG8). Similarly, professional astronomy and physics societies (e.g. the American Astronomical Society, European Astronomical Society, African Astronomical Society, Astronomical Society of India, etc.) play an important role in bringing together professionals at all career stages and advancing the field of astronomy and astroinformatics. These organizations also provide advice to governmental organizations and help foster collaborations and partnerships across the globe.

In many places, astroinformatics is not yet offered as a course of study and very few institutions offer an astronomy degree in Southeast Asia. Building curricula and faculty who can teach it takes time, and remote on-line courses could be a temporary solution for this challenge.

Like other interdisciplinary fields, the nature of astroinformatics leads to a struggle for its practitioners due to:
i) Lack of recognition of their work (lack of standard publication venues)
ii) Lack of system-wide recognized career achievements and rewards, etc.

Astroinformatics professionals are neither here nor there and are therefore not fully included and welcomed by the different fields they subtend. On the other hand, interdisciplinary work can be rewarding and if properly integrated at the university level can benefit from support from multiple departments. Successful stories of such collaboration should be shared widely within governments, funding agencies as well as universities.

*Regional and Global Collaborations*: Astronomy brings together advanced scientific research, state-of-the-art technology, and educational initiatives, all while captivating and stimulating people of all ages. By doing so, it possesses the potential to serve as a powerful catalyst for sustainable global development and the resolution of global societal issues. It attracts a diverse range of scientists and experts from various fields, fostering collaboration and innovation.

---

[4] Wilkinson, M. et al. The FAIR Guiding Principles for scientific data management and stewardship. Sci Data 3, 160018 (2016)
[5] For instance, in the United States, from 2018-2020, men earned 64% of the PhDs. This fraction is not dissimilar from the statistics reported by the International Astronomical Union whose membership is 78% male but the participation of female members increases in younger age groups.
[6] In the United States for example, the fraction of Black, Hispanic, or Native American astronomers is < 2% even though these groups make up ~30% of the US population. In South Africa, the number of black South Africans in astronomy is growing but remains small and highly disproportionate from the population.

The large amount of data obtained by modern astronomical observational research not only solves the mysteries of the universe but also serves as an intellectual property shared by humanity. Regional and global collaborations in astronomy present significant opportunities for advancing scientific understanding and promoting global development.

**Opportunities for Collaboration**

1. Encourage International Cooperation: Astronomy is one of the rare fields where a large fraction of science data is available for scientists to download without many barriers. Promote international cooperation and collaboration to help share resources and expertise, facilitate data sharing, and ensure that data is available to researchers worldwide. This enables the actual use of data—which is the primary aim of taking the data in the first place. To promote smooth collaboration through favourable policies and regulatory frameworks, governments must develop explicit agreements on key aspects like data sharing, intellectual property rights, technology transfer, and people-to-people exchanges. By doing so, they can ensure that all parties involved know their rights and responsibilities from the outset. Science 20 academies can assist their governments in promoting collaboration, exchange programs, and joint research initiatives by fostering diplomatic relations and scientific cooperation agreements with other nations.
2. Building Standards: Astronomy has taken important strides towards the development, use and adoption of standardized data formats and metadata, to help facilitate data sharing and interoperability among different data archives and scientific communities. Ensure that this continues and facilitate the building of tools that incorporate the standards. As part of an Open Science approach, the algorithms and software supporting machine learning, other analytics, and models, should be made available and be well-documented to support reproducibility of results.
3. Data Provenance: To ensure comprehensive traceability, it is important to keep track of input and output, and configuration parameters, as well as the reference to the software used, along with its version. This approach allows for a complete record of all relevant information related to the analysis process. This process can be divided into two main components: the technical pipeline and the human-oriented aspect. Educate and inculcate best practices for this. This will also help with ethical use of data and reproducibility of results in astronomy and in other fields.
4. Data sharing: Invest in the development and maintenance of robust data storage infrastructure to ensure the long-term preservation and accessibility of data. Open-access databases, and platforms facilitate the sharing and dissemination of research findings and datasets in an equitable way. In addition, governments can support initiatives that ensure responsible use of shared data. To achieve this, they can establish data-sharing policies and protocols that address privacy and security concerns, implement secure data management systems following cybersecurity best practices, and develop guidelines for the responsible use of shared data. By taking these steps, Science 20 academies can assist their governments in fostering a culture of open collaboration and maximise the scientific potential of astronomical research.
5. Gap Analysis and Sustainability: Science 20 academies can assist their governments to address resource gaps and ensure the long-term sustainability of collaborative projects. Collaboration among funding agencies should be encouraged to develop joint funding opportunities that specifically support international projects in the field of astronomy.
6. Education and Outreach: Promote the use of survey data in education and outreach activities to engage the public and inspire the next generation of astronomers and data scientists. In this context it is important that astronomical survey data are made publicly available (which is already the case for many surveys) for use by non-scientists, educators, and outreach professionals. This offers a unique opportunity to engage communities in countries with less direct access to astronomical survey facilities.
7. Capacity Building and Training: Through capacity-building programs, G20 governments should bolster scientific expertise and infrastructure in countries with limited resources. To achieve this, they can offer scholarships, grants, and training opportunities for scientists from developing regions, enabling them to collaborate with leading astronomical institutions within G20 countries and acquire valuable skills and knowledge. By fostering collaboration in training and capacity-building initiatives, facilitating the exchange of knowledge and expertise, developing mentorship programs, and encouraging scientific exchanges, governments can promote skill development and facilitate technology transfer among partners.
8. Cross-disciplinary Research: Encourage cross-disciplinary research by providing opportunities for astronomers to collaborate with researchers from other fields, such as computer science and statistics, but also fields such as geography and medicine, and sub-disciplines (other wavelengths and messengers), to explore new methods of data analysis and visualization.
9. Recognition and Rewards: The astronomical community now expects survey data to be publicly available in a timely fashion which means that data rights for surveys can only be exercised for a limited period. This can be demotivating for (in particular young) researchers who want to contribute to surveys but also want to benefit from proprietary periods to create and publish scientific results, which are still an important component in evaluations for stable jobs in academia. The academic community, with support from governments, should investigate ways of creating career paths where one is rewarded and promoted for one's contribution to making large surveys possible. This requires the development of evaluation criteria that balance individual versus team efforts and reward contributions to open science.
10. Inclusivity and Diversity: An inclusive environment is critical for a diverse workforce to thrive.
    a. Outreach towards historically excluded and/or underrepresented groups should consider their needs and barriers.
    b. For retention, cluster recruitment and hiring, dual mentoring models, mentor pod models should be regularly employed, avoiding the mentor-mentee and the master-apprentice models.
    c. Paying fair wages and imparting knowledge of all aspects of finances to early career students, especially those from impoverished backgrounds is critical.
11. Programs for Identification of Hidden Biases: It is also important to train those who are *not* from historically excluded groups to learn how to recognize their own bias and actively work against such biases. Experiences with double blind reviewing processes at funding agencies, and in various time allocation committees have shown increases in the diversity of successful proposers, particularly increasing the number of first-time proposers and early career researchers. However dual anonymous or double-blind reviews cannot fundamentally change the demographics of a field and that requires more active efforts such as those described above.

**References**
1. Pathways to Discovery in Astronomy and Astrophysics for the 2020s, Decadal Survey on Astronomy and Astrophysics 2020 (2021)
2. The Astronet Science Vision and Infrastructure Roadmap 2022–2035 (2023)5

**S20 Co-Chair**: Ashutosh Sharma, Indian National Science Academy
**INSA S20 Coordination Chair:** Narinder Mehra, Indian National Science Academy
**Contributors:**
Ashish Mahabal, California Institute of Technology, USA
Pranav Sharma, Indian National Science Academy, India
Rana Adhikari, California Institute of Technology, USA
Mark Allen, Centre de Données astronomiques de Strasbourg, France
Stefano Andreon, INAF-OA Brera, Italy
Varun Bhalerao, Indian Institute of Technology - Mumbai, India
Federica Bianco, University of Delaware, USA
Anthony Brown, Leiden University, Netherlands
S. Bradley Cenko, NASA GSFC, USA
Paula Coehlo, University of São Paulo, Brazil
Jeffery Cooke, Swinburne University of Technology, Australia
Daniel Crichton, Jet Propulsion Laboratory - NASA, USA
Chenzhou Cui, National Astronomical Observatories, China
Reinaldo de Carvalho, Universidade Cidade de São Paulo, Brazil
Richard Doyle, Jet Propulsion Laboratory - NASA, USA
Laurent Eyer, University of Geneva, Switzerland
Bernard Fanaroff, South African Radio Astronomy Observatory, South Africa
Christopher Fluke, Swinburne University of Technology, Australia
Francisco Forster, Universidad de Chile, Chile
Kevin Govender, Office of Astronomy for Development, South Africa
Matthew J. Graham, California Institute of Technology, USA
Renée Hložek, University of Toronto, Canada
Puji Irawati, National Astronomical Research Institute, Thailand
Ajit Kembhavi, Inter-University Centre for Astronomy and Astrophysics, India
Juna Kollmeier, Canadian Institute for Theoretical Astrophysics, Canada
Alberto Krone-Martins, University of California Irvine, USA
Shri Kulkarni, California Institute of Technology, USA
Giuseppe Longo, Università degli Studi di Napoli Federico II, Italy
Vanessa McBride, Office of Astronomy for Development, South Africa
Jess McIver, University of British Columbia, Canada
Sanjit Mitra, Inter-University Centre for Astronomy and Astrophysics, India
Timo Prusti, European Space Agency (ESA), Netherlands
A. N. Ramaprakash, Inter-University Centre for Astronomy and Astrophysics, India
Eswar Reddy, Indian Institute of Astrophysics, India
David H. Reitze, California Institute of Technology, USA
Reinaldo R. Rosa, Instituto Nacional de Pesquisas Espaciais-INPE, Brazil
Rafael Santos, Instituto Nacional de Pesquisas Espaciais, Brazil
Kazuhiro Sekiguchi, National Astronomical Observatory of Japan, Japan
Kartik Sheth, NASA, USA
Seetha Somasundaram, India Space Research Organization, India
Tarun Souradeep, Raman Research Institute, India
R. Srianand, Inter-University Centre for Astronomy and Astrophysics, India
Annapurni Subramaniam, Indian Institute of Astrophysics, India
Alex Szalay, Johns Hopkins University, USA
Shriharsh Tendulkar, Tata Institute of Fundamental Research, India
Laura Trouille, The Adler Planetarium, USA
Yogesh Wadadekar, National Center for Radio Astrophysics - TIFR, India
Patricia Whitelock, South African Astronomical Observatory, South Africa